\documentclass[10pt,aps,pre,twocolumn,
superscriptaddress, nofootinbib, notitlepage]{revtex4-1}

\usepackage{amssymb,amsmath}
\usepackage{natbib}
\usepackage[bb=boondox]{mathalfa}
\usepackage{graphicx}
\usepackage{amsthm}
\usepackage{cancel}
\usepackage{color}
\usepackage{bbold}
\usepackage{wrapfig}
\usepackage{booktabs}
\usepackage{float}
\usepackage{enumitem}
\usepackage{hyperref}
\usepackage{tikz}
\usepackage{comment}
\usepackage{longtable}
\usepackage{multirow}

\newtheorem{obs}{Observation}

\begin{document}

\title{Traffic flow splitting from crowdsourced digital route choice support}

\author{David-Maximilian Storch}
\affiliation{Chair for Network Dynamics, Center for Advancing Electronics Dresden (cfaed), Technical University of Dresden, 01062 Dresden, Germany}
\affiliation{Institute for Theoretical Physics, Technical University of Dresden, 01062 Dresden, Germany}
\author{Malte Schr\"oder}
\affiliation{Chair for Network Dynamics, Center for Advancing Electronics Dresden (cfaed), Technical University of Dresden, 01062 Dresden, Germany}
\affiliation{Institute for Theoretical Physics, Technical University of Dresden, 01062 Dresden, Germany}
\author{Marc Timme}
\affiliation{Chair for Network Dynamics, Center for Advancing Electronics Dresden (cfaed), Technical University of Dresden, 01062 Dresden, Germany}
\affiliation{Institute for Theoretical Physics, Technical University of Dresden, 01062 Dresden, Germany}
\affiliation{Lakeside Labs, Lakeside B04b, 9020 Klagenfurt, Austria}
\date{\today}


\begin{abstract}
Digital technology is fundamentally transforming human mobility. Route choices in particular are greatly affected by the availability of traffic data, increased connectivity of data sources and cheap access to computational resources. Digital routing technologies promise more efficient route choices for the individual and a reduction of congestion for cities. Yet, it is unclear how widespread adoption of such technologies actually alters the collective traffic flow dynamics on complex street networks. Here, we answer this question for the dynamics of urban commuting under digital route choice support. Building on the class of congestion games we study the evolution of commuting behavior as a fraction of the population relies on, but also contributes to, crowdsourced traffic information. The remainder of the population makes their route choices based on personal experience. We show how digital route choice support may cause a separation of commuter flows into technology and non-technology users along different routes. This collective behavior may fuel systemic inefficiencies and lead to an increase of congestion as a consequence. These results highlight new research directions in the field of algorithmic design of route choice decision support protocols to help fight congestion, emissions and other systemic inefficiencies in the course of increasing urbanization and digitization.
\end{abstract}

\maketitle

\section{Introduction}

Within the past decade, the smartphone has become a daily companion to as many as 2.71 billion people around the globe \cite{Statista2019}. Its vast supply of mobile applications offers software solutions that connect people, provide information and simplify many tasks of every-day life. 

In mobility in particular, the smartphone has become an indispensable tool that reshapes the way people get from $A$ to $B$ \cite{Statista2013,Statista2017,Batty2012,Shaheen2016}: It enables new forms of mobility services (e.g. ride-sharing, e-hailing, car-sharing) \cite{Furuhata2013,Santi2014,Tachet2017,Vazifeh2018,Molkenthin2019}, easy and transparent access to information (e.g. public transport schedules, street maps, real-time traffic updates) \cite{Weissmann2012}, as well as decision support in situations under uncertainty (e.g. route choice recommendations) \cite{Essen2016}.

Despite the widespread adoption of smartphone technology for mobility services, little is known theoretically about the mechanisms of how phone-based mobility applications modify the collective dynamics of complex mobility systems \cite{Acemoglu2018,Shaheen2016}. Here, we focus on mobile applications offering route choice recommendations in traffic situations under uncertainty \cite{Daganzo08}: In those situations, route choice applications provide their users with valuable traffic updates (otherwise unavailable to them) and advise them on how to navigate the local environment (e.g. described by a street network) ideally to minimize expected travel times \cite{Bertsekas1991}, volatility \cite{Levinson2003}, or environmental harm \cite{Ericsson2006,MeinGruen2019}. 

Research on route choice behavior has helped to characterize the impact of selfish routing on travel times and the emergence of congestion \cite{Braess1968,Roughgarden2002,Roughgarden2003,Roughgarden2007}. Recent availability of massive amounts of traffic data has enabled the investigation of traffic assignment and network flows in real cities to complement theoretical efficiency bounds \cite{Youn2008,Colak2016}.

It is easy to imagine that app-based routing information as well as smart mode and route suggestions may help to overcome the negative effects of selfish routing and benefit all traffic participants. Even without dedicated measurements, crowdsourced information may provide a very accurate picture of the current and average traffic conditions \cite{OKeeffe2019}. Yet, studies on mobility systems with digital route choice support have mostly focused on their potential efficiency, technical implementation and their computational performance \cite{Phuttharak2019} with limited emphasis on the induced collective traffic assignment dynamics \cite{Shaheen2016,Essen2016}. 
However, given the complex interplay of route suggestions, actual route choices and street network, it is far from clear how the global state of traffic actually responds to widespread use of digital route choice support.

\begin{figure*}
\centering
\includegraphics[width=\linewidth]{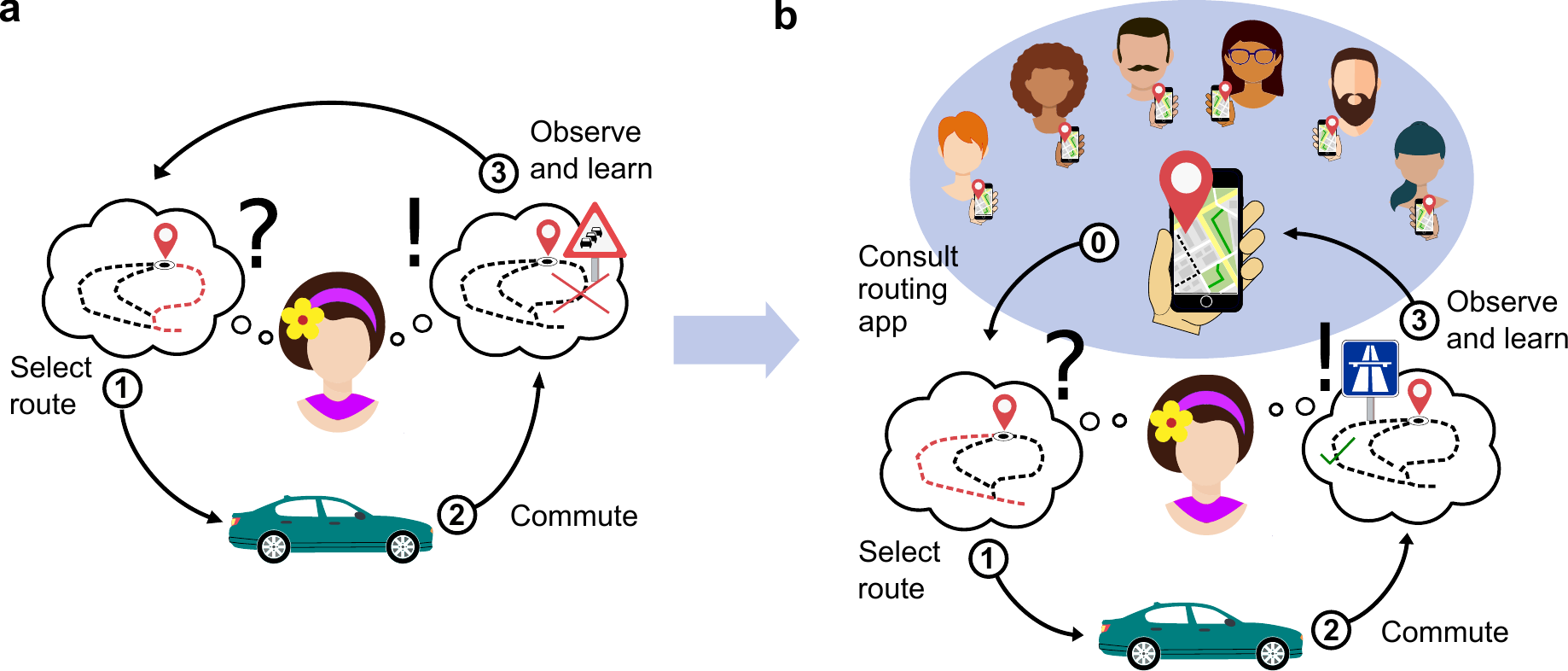}

    \caption{\textbf{Digital decision support changes route choice behavior.} \textbf{a}, Without digital decision support commuters only rely on their personal learning about the performance of different routing options. In a daily repeated cycle, they \textit{select a route} (1), \textit{commute} to their destinations (2), \textit{observe and learn} the travel time along their chosen routes (3). Their observations and learnings serve as a foundation for future route choices. \textbf{b}, With digital decision support commuters integrate crowdsourced traffic information from their fellow travelers (shading) on top of their personal learning. Before embarking on a trip, commuters \textit{consult their routing apps} for route choice advise (0), \textit{select a route} conditional on the information obtained and their personal experience (1), \textit{commute} to their destinations (2), \textit{observe and learn} which routes perform best (3). Fully automated, commuters share information on their latest travel times with the digital platform provider (crowdsourcing) to update and improve the quality of the recommendations of the digital decision support tool \cite{data_Image}.}
\label{fig:FIG1_DecisionSupport}
\end{figure*}

In this article, we address this issue for the dynamics of urban commuting. We propose a reformulation of congestion games \cite{Rosenthal1973,Colak2016} driven by an adaptive payoff-based learning rule \cite{Cominetti2010} where heterogeneous population shares subscribe to a digital route choice support service, while others rely on personal commuting experience. We integrate a digital platform operator who provides mode and route choice recommendations based on traffic information crowdsourced from his subscribers (see Fig.~\ref{fig:FIG1_DecisionSupport}). Commuters balance between following his advice, or relying on their personal experience in making a route choice.

We demonstrate the emergence of flow separation into groups of technology and non-technology users along different routes of the street network, stimulated by the collective sharing of crowdsourced traffic information. Furthermore, we highlight how flow splitting may contribute to inefficient use of available street capacity and increase the overall travel time for the population. 

Our results promise new insights into technology-enabled traffic assignment on complex mobility systems. Not only do the results show that informational nudges influence individual route choices, but also how they impact the overall state of traffic. As Mobility-as-a-Service platforms become more and more common, our results highlight further research directions in designing route choice support protocols that reduce congestion, emissions or other forms of systemic inefficiencies, instead of fueling them. 

\section{Urban commuting under digital route choice support}

In the following, we introduce the basic notation for the congestion game used to analyze the commuting dynamics in urban environments under digital route choice support. First, we define the urban environment that is inhabited by two groups of commuters -- technology users and non-technology users -- who have a daily recurring commuting demand to travel between fixed origin-destination pairs. Second, we present the commuters' route choice problem and show how digital decision support modifies the route choice dynamics. Third, we detail how a digital platform operator crowdsources traffic observations to provide commuters with smart mode and route choice recommendations.

\subsection{Urban environment}
\label{sec:UrbanEnvironment}

We formalize the urban commuting dynamics in terms of flows on a network. A graph defines the urban street network which flows materialize on. A set of latency functions governs the street network's resilience to congestion.

Consider the street network to be described by the graph $G=(\mathcal{V},\mathcal{E})$, whose edges $\mathcal{E}=\{1,2,\dots,E\}$ denote street segments and nodes $\mathcal{V}=\{1,2,\dots,V\}$ correspond to intersections. Assume streets $e\in\mathcal{E}$ to be common pool resources \cite{Ostrom1990} which are equipped with a latency function $c_e:\mathbb{N}\rightarrow\mathbb{R}$ each. The latency function determines the duration of travel along the street segment. It is a monotonously increasing function of the vehicle flow $f_e\in\mathbb{N}$, and formalizes the street segment's resilience to congestion \cite{Ni2016}. Common edge latency functions are of polynomial form \cite{Branston1976}, e.g.
\begin{align}
    c_e(f_e)=c_0\left(1+B_e\left(\frac{f_e}{f_0} \right)^4\right),
\end{align}
where $c_0$ is the free flow latency, $f_0\in \mathbb{N}$ denotes an effective capacity that governs the onset of congestion and $B_e\geq 0$. If the product $c_0B_e\to 0$ an edge reduces to a common good that is non-rivalrous in consumption and resilient to congestion. Else, the edge's latency rises substantially as $f_e$ approaches or exceeds $f_0$.

In this setting, distribute an urban population $\mathcal{N}=\{1,2,\dots,N\}$ on $G$, and assume that each individual $i\in\mathcal{N}$ has a daily commuting demand $D_i=\{o,d\}$ to travel from some origin $o\in\mathcal{V}$ to some destination $d\in \mathcal{V}$. Without loss of generality, we require $o\neq d$. 

\subsection{Commuter perspective}

Every day, the population participates in a congestion game. Each commuter $i\in\mathcal{N}$ tries to find an ideal route that minimizes her personal travel time while accommodating her mobility demand $D_i$ \cite{Wardrop1952,Rosenthal1973}. Therefore, commuters choose one of $S_i$ routes from their finite strategy profiles $\mathcal{S}_i=\{p_i^1,p_i^2,\dots, p_i^{S_i}\}$. The elements of $\mathcal{S}_i$ correspond to directed paths on $G$ connecting the origin-destination pair $D_i$ and may be partially overlapping. 

Each routing option $p_i^s\in\mathcal{S}_i$ yields a travel time $\Phi_{i^s}$, defined in terms of the total latency $\sum_{e\in p_i^s} c_e$ along the path corresponding to individual $i$'s strategy $s\in\{1,2,\dots,S_i\}$. However, at the time of decision-making commuters typically do not know about the edge flows $f_e, e\in\mathcal{E}$, and latencies to be realized that day. For them, the quantities are random variables, rendering it highly non-trivial to anticipate which route will be the fastest. Some routes may seem short in terms of distance, but may turn out to be heavily congested. Others might seem long, but may be less utilized by the population. Hence, commuters face a decision problem on how to navigate an urban environment with stochastic latencies most efficiently. 

\subsubsection{Strategic interaction} 
Commuters interact through increased travel times when they use the same streets. Beyond these direct, traffic-mediated interactions between commuters, technology-based route choice support introduces additional indirect interactions, for example due to collective sharing of crowdsourced traffic information. These higher-order interactions also affect non-technology users through their direct interaction with technology users.

\begin{figure}
    \centering
    \includegraphics{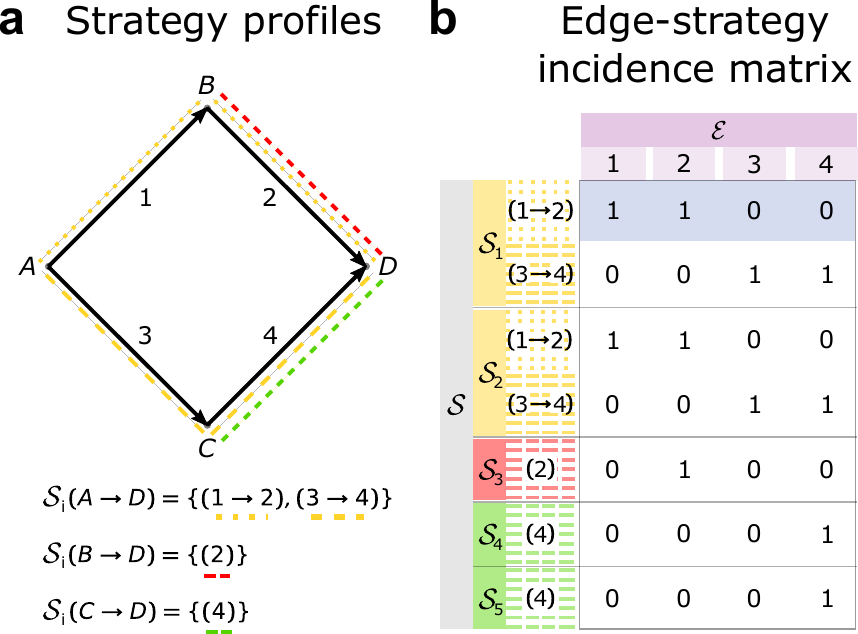}
    \caption{\textbf{Formalizing route choice decisions.} The street network's topology, latency structure and the origin-destination distribution define a strategic interaction between commuters. \textbf{a}, Strategy profiles for three exemplary OD-pairs ($\{A\to D,B\to D,C\to D\}$) in a four-edge graph. The first strategy profile contains two routing strategies while the other two contain a single option only. \textbf{b}, Transpose of the edge-strategy incidence matrix $I^\mathsf{T}$ for five exemplary commuters $i\in\{1,2,\dots,5\}$. For example, commuter 1 needs to travel from $A$ to $D$. Hence, she has two strategies in her strategy profile $\mathcal{S}_1$ (yellow). Her first strategy (dotted) includes travel along edges $\{1,2\}\subset\mathcal{E}$, denoted as ones in the edge-strategy incidence matrix (blue shading), but excludes other edges, denoted as zeros. Her second strategy (dashed) follows edges $\{3,4\}\subset\mathcal{E}$, corresponding to ones in the third and fourth column of the edge-strategy incidence matrix, but no non-zero entries otherwise. Note that the incidence matrix duplicates paths on $G$ as different strategies if they correspond to different individuals.}
    \label{fig:FIG2_Nomenclature}
\end{figure}

Independent of technology support, direct interactions between commuters can be classified according to three categories, depending on their mutual strategic routing options. The overlap in routing options introduces a strategic interaction between the participants of the urban commuting game, caused by the rivalry for low latency street segments which are non-excludable.
\begin{itemize}
    \item \textit{Identical strategy profiles.} Identical origin-destination pairs $D_i=D_j, \ i,j\in\mathcal{N}$, have the same routing options to complete their daily trips such that the strategies $p_i^s$ and $p_j^s$ correspond to the same path on $G$. Hence, identical origin-destination pairs may be envisaged as identical species of commuter profiles that are likely to travel along the same routes. When doing so, they increase each others' travel times.
    \item \textit{Intersecting strategy profiles.} Commuters with different origin-destination pairs $D_i\neq D_j, \ i,j\in\mathcal{N}$, might still have significant overlap in terms of joint street segments along the different routing options in their strategy profiles. Depending on their respective route choices, they might meet along selected street segments and influence each other's travel times. 
    \item \textit{Disjoint strategy profiles.} In contrast, different origin-destination pairs $D_i\neq D_j, \ i,j\in\mathcal{N}$, can also imply disjoint strategy profiles. In this case, commuters do not meet along the streets on $G$. Their respective travel times are not directly dependent on one another.
\end{itemize}
The congestion game may be envisaged as a complex many-body system consisting of heterogeneous commuter species who are defined by their unique sets of strategy profiles. Depending on the degree of overlap between their strategies, the system mediates a strong repulsive coupling between the different specimen. 

Typically, commuters have no knowledge about which person refers to which of the three categories. They may only infer information on the statistical properties of the origin-destination distribution based on their travel time observations.\\ 

\paragraph{Edge-strategy incidence matrix.} To formalize the information on strategies and graph topology, we define the edge-strategy incidence matrix $I$. It is a linear map from the joint space of strategies $\mathcal{S}=\cup_{i\in\mathcal{N}} \mathcal{S}_i=\{p_1^1,\dots, p_1^{S_1},p_2^1,\dots,p_N^{S_N}\}$ to edge space $\mathcal{E}$, defined by 
\begin{align}
    I_{e,i^s} = \begin{cases} 
                1, \quad \mathrm{if}\ e\in p_i^s,\ p_i^s\in\mathcal{S}\\
                0, \quad \mathrm{otherwise.}
              \end{cases}
              \label{eqn:EdgeStrategyMatrix}
\end{align}
Fig.~\ref{fig:FIG2_Nomenclature} shows an illustrative example of how $I$ is constructed from both $\mathcal{E}$ and $\mathcal{S}$.

To investigate the day-to-day dynamics of urban commuting we consider constant $I$, i.e. stationary graph topology and origin-destination demands since the commuting dynamics happens on much faster timescales. Typically, the graph's topology or commuting demand patterns evolve on timescales of months or years (e.g. upon infrastructural reconstruction in the urban area or urbanizaton effects such as im-, emigration or sprawl), thereby only slowing modulating $G$ and $\mathcal{S}$ compared to the day-to-day level. 

\subsubsection{Route choice} 

Denote by $a_i^s=1$ person $i$'s choice for route $p_i^s\in \mathcal{S}_i$, implying a decision against her strategic alternatives, namely $a_i^r=0$ for all $r\in \mathcal{S}_i\backslash\{p_i^s\}$. For each person $i\in \mathcal{N}$, let $\mathbf{e}_i^{s}$ be the unit vector pointing in the direction of her route $s$ within the joint strategy space of all players $\mathcal{S}$. Denote the population's collective route choice $\mathbf{a}\in \{0,1\}^{|\mathcal{S}|}$ by
\begin{align}
\mathbf{a} =\sum_{i\in\mathcal{N}}\mathbf{a}_i = \sum_{i\in\mathcal{N}} \sum_{s=1}^{S_i} a_i^s \mathbf{e}_i^s,
\end{align}
with one entry of $\mathbf{a}_i$ being one and all others being zero.
Note that even players traveling along identical routes will have different $\mathbf{a}_i$ since $\mathbf{e}_i^s$ points towards a player-specific strategy on the joint space of all strategies $\mathcal{S}$.

Collective route choices $\mathbf{a}$ fully determine the system state and translate into edge flows
\begin{align}
    \mathbf{f}=\sum_{e\in\mathcal{E}} f_e \mathbf{e}_e = I\mathbf{a},
\end{align}
via the edge-strategy incidence matrix $I$. Here, $\mathbf{e}_e$ denotes the unit vector pointing in the direction of edge $e\in\mathcal{E}$.

Let $C:\mathbb{N}^{|\mathcal{E}|}\rightarrow \mathbb{R}^{|\mathcal{E}|}$ be the operator mapping edge flows to latencies via
\begin{align}
    C(\mathbf{f})=\sum_{e\in\mathcal{E}} c_e(f_e) \cdot \mathbf{e}_e.     
\end{align}
It defines the trip duration performance map 
\begin{align}
    \Phi(\mathbf{a}) = I^\mathsf{T} C(I\mathbf{a}) \in \mathbb{R}^{|\mathcal{S}|}
    \label{eqn:PayoffPerformanceMap}
\end{align}
which assigns a travel time to each routing option $p_i^s\in \mathcal{S},\ i\in\mathcal{N}$ for given collective action $\mathbf{a}$.

\subsubsection{Adaptive payoff-based learning dynamics without digital route choice support}

Commuters use the trip duration performance map Eqn.~\eqref{eqn:PayoffPerformanceMap} to determine an optimal route choice $\mathbf{a}_{i,t}$ on day $t$. However, they do not know how the other commuters will decide that day ($\mathbf{a}_{-i,t}=\sum_{\substack{j\in\mathcal{N}\backslash \{i\}}} \mathbf{a}_{j,t}$), nor do they have information on the others' strategy profiles to infer a best response. Hence, $\mathbf{a}_t$ is a random variable to them, and so is the travel time of their routing options. To deal with this uncertainty, commuters form beliefs about the expected trip duration of their strategies based on past experience. Thereby, they learn which of their routing options, on average, are best aligned with their objective to reduce commuting time. The evolution of travel time beliefs follows an adaptive payoff-based learning rule \cite{Cominetti2010}, which consists of an iterated three-step cycle:
\begin{enumerate}
    \item \textit{Select route}: Commuters choose a route from their strategy profiles based on their personal travel time beliefs from past observations.
    \item \textit{Commute}: Commuters travel to their final destination along the chosen route for a duration determined by the travel time performance map.
    \item \textit{Observe and learn}: Commuters update the travel time belief corresponding to their route choice.
\end{enumerate}
Fig.~\ref{fig:FIG1_DecisionSupport}a shows an illustrative visualization of this timing.\\

1. \textit{Select route.} Denote by $x_{i,t}^s$ commuter $i$'s travel time belief for $\Phi(\mathbf{a}_t)\cdot \mathbf{e}_i^s$ for route $s\in\{1,2,\dots,S_i\}$ before embarking on their trip on date $t$. The population's joint belief vector can be written as
\begin{align}
    \mathbf{x}_t = \sum_{i\in\mathcal{N}}\mathbf{x}_{i,t} =  \sum_{i\in\mathcal{N}}\sum_{s=1}^{S_i} x_{i,t}^s \cdot \mathbf{e}_i^s.
\end{align}
It evolves in time and governs the realization of the collective route choice. In line with the human choice model \cite{McFadden1975,Ben-Akiva1985}, beliefs $\mathbf{x}_t$ at time $t$ translate into commuter-specific mixed strategy profiles 
\begin{align}
    \boldsymbol{\pi}_{i,t}=\sum_{s=1}^{S_i}\pi_{i,t}^s \mathbf{e}_i^s= \sum_{s=1}^{S_i} \frac{e^{-\beta_i x_{i,t}^s}}{\sum_{r=1}^{S_i} e^{-\beta_i x_{i,t}^r}} \mathbf{e}_i^s
\end{align}
via a Boltzmann distribution. Here, $\pi_{i,t}^s$ denotes the probability for $i$ to select route $s\in\{1,2,\dots,S_i\}$. $\beta_i\in\mathbb{R}_0^+$ interpolates between purely opportunistic ($\beta_i\to\infty$) and uniformly random ($\beta_i\to 0$) behavior.

The joint population's mixed strategy profile is defined by
\begin{align}
\boldsymbol{\pi}_t=\sum_{i\in\mathcal{N}} \boldsymbol{\pi}_{i,t}.
\end{align}\\

2. \textit{Commute.} Each day $t$, all commuters make their route choice for an $s\in\{1,2,\dots,S_i\}$ by realizing the random variable $\mathbf{a}_{i,t}$ according to $\boldsymbol{\pi}_{i,t}$. Afterwards they start to travel along the chosen route from their respective origin to their destination. The flows $\mathbf{f}$ materialize instantaneously along the edges of $G$.\\

3. \textit{Observe and learn.} Commuters observe travel times as specified in Eqn.~\eqref{eqn:PayoffPerformanceMap}. Upon completion of the trip, they compare $\mathbf{x}_t$ and $\Phi(\mathbf{a}_t)$ and update their beliefs according to
\begin{align}
    \mathbf{x}_{t+1}-\mathbf{x}_t &=  \mathrm{diag}\left(\mathbf{a}_t \right)\left[ \Phi(\mathbf{a}_t)-\mathbf{x}_t \right]\\
    &= \mathrm{diag}\left(\mathbf{a}_t \right)\left[ I^\mathsf{T} C\left[I\mathbf{a}_t\right]-\mathbf{x}_t \right]. \label{eqn:DynamicalSystem1}
\end{align}
This mimics an iterated learning process, where discrepancies $\Phi(\mathbf{a}_t)-\mathbf{x}_t$ between beliefs $\mathbf{x}_{t}$ and actual travel times $\Phi(\mathbf{a}_t)$ that day induce a belief correction. However, commuters will only update the beliefs for the routes that they have actually traveled along and where they were able to make a travel time observation [projector $\mathrm{diag}(\mathbf{a}_t)$ in Eqn.~\eqref{eqn:DynamicalSystem1}]. The resulting prior belief in game stage $t+1$ modifies the mixed strategy profile and the commuting game repeats itself. 

\subsubsection{Digital route choice support} 

The widespread availability of smartphone-based route choice support applications modifies the adaptive payoff-based learning dynamics from Eqn. \eqref{eqn:DynamicalSystem1}, and defines an indirect interaction. The applications provide commuters with traffic information and thereby modify their users' mixed strategy profiles. Besides updating travel time beliefs solely based on personal observations, commuters increasingly rely on additional traffic updates from their apps before embarking on their daily commute. Such updates typically come in the form of travel time performance signals $\boldsymbol{\tau}_t\in\mathbb{R}^{|\mathcal{S}|}$ along different routing options on $G$. The signals enable their receivers to make a more informed decision on how to navigate the urban traffic most effectively. Informational signals induce an additional belief update (\textit{consult routing app}) in the decision cycle shown in Fig.~\ref{fig:FIG1_DecisionSupport}b:
\begin{enumerate}
    \item[0.] \textit{Consult routing app}: Commuters consult a digital decision support tool prior to making a route choice to calibrate their travel time beliefs.\\
\end{enumerate}
The subsequent steps 1-3 are identical to pure payoff-based adaptive learning explained in the previous section, apart from sharing personal travel time information with the digital platform provider in step 3 (crowdsourcing). The commuters' additional belief update from consulting the mobile application modifies the formation of mixed strategy profiles $\boldsymbol{\pi}_t$ and impacts all subsequent steps:\\

0. \textit{Consult routing app.} Assume commuter $i\in\mathcal{N}$ uses a route choice app in her day-to-day commute. Her travel time belief at time $t$ is defined by Eqn.~\eqref{eqn:DynamicalSystem1}. It reflects her last commuting experiences. To benefit from the app's capability to monitor the statistical performance of large sets of routes at the same time, she tests how well her personal beliefs are calibrated with the signal $\boldsymbol{\tau}_{i,t}$ she receives. She determines the difference between the two, 
\begin{equation*}
   \boldsymbol{\tau}_{i,t}-\mathbf{x}_{i,t} ,
\end{equation*}
and updates her travel time beliefs accordingly discounted by the scalar parameter $\kappa_i\in [0,1]$. $\kappa_i$ determines to what extent a person conditions her beliefs on the mobile applications' signal, or her personal experience. $\kappa_i=0$ implies that she ignores the tools' input, while $\kappa_i=1$ indicates full replacement of personal observations by the tools' recommendations. Empirical evidence suggests significant trust, i.e. large $\kappa_i$ \cite{Robinette2016,Milner2016}.

\begin{widetext}
The learning dynamics under digital decision support becomes a superposition of both the personal experience contribution, as well as the integration of the decision support tools' signals
\begin{align}
    \mathbf{x}_{t+1}-\mathbf{x}_t = (1-\kappa)\ \mathrm{diag}(\mathbf{a}_t) \left[ \Phi(\mathbf{a}_t)-\mathbf{x}_t \right] + \kappa \left[ \boldsymbol{\tau}_{t+1}-\mathbf{x}_t\right].
    \label{eqn:DecisionSupportDynamics}
\end{align}
In this notation,  $\kappa=\sum_{i\in\mathcal{N}} \sum_{s=1}^{S_i}  \kappa_i\cdot  \mathbf{e}_i^{s^\mathsf{T}} \mathbf{e}_i^s$ is the diagonal matrix whose entries bundle the individual trust parameters of the different commuters. 
Note that setting $\kappa_i = 0$ corresponds to not using the decision support tool for the corresponding individual. Therefore, it becomes possible to study the impact of heterogeneous groups of technology and non-technology users on the urban commuting dynamics by composing the $\kappa$-matrix accordingly (as we will detail in the following).
\end{widetext}

\subsection{Digital platform operator perspective}

Let there be a digital platform operator who curates the digital route choice support tool that provides the population with route choice recommendations. The tool is built around a statistical model on the latency of the mobility system that is continuously being updated to generate better traffic forecasts. Commuters can request route choice recommendations based on this model free of charge, but they agree to provide the platform operator with spatiotemporal movement data on their commuting behavior and travel time performance every time they utilize the routing service. This process of data aggregation is called \textit{crowdsourcing} \cite{Chatzimilioudis2012} and enables the platform provider to collect massive amounts of traffic observations in a single game stage to improve his statistical model (in contrast to individual commuters who learn based on their private observations only).\\

\paragraph{Crowdsourcing.} Let $\mathcal{T}\subset \mathcal{N}, |\mathcal{T}|=T,$ be the subset of the commuter population that consults their smartphone applications before a route choice ($\kappa_i \geq 0$ for all $i \in \mathcal{T}$). The remainder of the population, $\mathcal{A}=\mathcal{N}\backslash\mathcal{T}, |\mathcal{A}|=A$, does not apply any technology support ($\kappa_i = 0, i \in \mathcal{A}$). $\mathcal{T}$ shares their individual travel time observations with the digital platform operator. That means, the provider receives information about the subset of edge flow realizations,
\begin{align}
    \mathbf{f}^T_t=R_t \mathbf{f}_t,
\end{align}
that technology users have traveled along on day $t$. Here, ${R}_t$ denotes the flow projector that projects onto the subset of edges in $\mathcal{E}$ which have been traveled along by members within $\mathcal{T}$. Note that $R_t$ is constructed from the joint action vector of technology users and thus evolves as a random variable in time, too. 
Depending on the composition of origin-destination pairs within $\mathcal{T}$, a small group of technology users is typically sufficient to crowdsource flow data about a large fraction of edges in $\mathcal{E}$ \cite{OKeeffe2019}.\\

\begin{figure*}
    \centering
    \includegraphics[width=\linewidth]{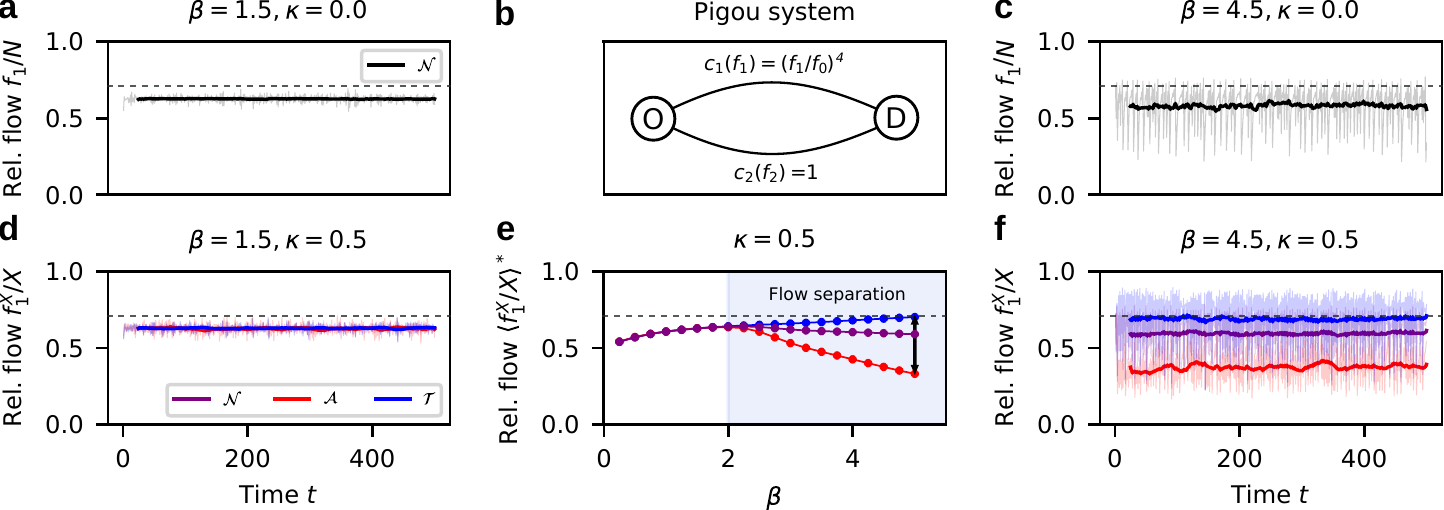}
    \caption{\textbf{Crowdsourced traffic information causes flow separation.} \textbf{b}, In the Pigou network, consisting of a single OD-pair connected by two routes, the population's traffic assignment changes from the state without decision support (top) to the state with decision support (bottom), where two qualitatively different regimes exist. \textbf{a, d}, For $\beta<\beta_\mathrm{crit.}$ technology and non-technology users equally contribute to the relative commuter flow (flow normalized by the respective group size) of the different routes. Qualitatively, the flow dynamics corresponds to the situation without digital decision support. \textbf{c, f}, For $\beta>\beta_\mathrm{crit.}$ the total flow of commuters ($\mathcal{N}$) splits into technology ($\mathcal{T}$) and non-technology users ($\mathcal{A}$). Technology users effectively dominate route 1 and displace non-technology users onto route 2. In the absence of digital decision support no such behavior can be observed. Bold lines depict moving averages of 25 time steps window size. \textbf{e}, The flow separation persists in user equilibrium and intensifies for larger values of $\beta$. Simulation parameters are $N=1000, T=700, f_0=700, \alpha=0.5$ (see Supplementary Material for details).}
    \label{fig:FIG3_PigouSystem}
\end{figure*}

\paragraph{Signal construction.} The platform operator acts as a social planner whose strategy it is to develop a refined statistical model of the typical edge latency on $G$. He updates the model continuously based on the crowdsourced data received by his subscribers $\mathcal{T}$. Similar to the learning dynamics assumed for commuters, let this model be within the class of exponential moving averages. That means, it is a weighted superposition of latest data points $\mathbf{f}_t^T$ at time $t$ and old observations $\mathbf{f}_{t-k}^T, k\in\{1,2,\dots,K\}$,
\begin{align}
   \boldsymbol{\tau}_{t+1}&= \alpha \Phi (\mathbf{f}_t^T) + (1-\alpha) \boldsymbol{\tau}_t \\
   &=    \alpha I^\mathsf{T} \sum_{k=0}^{K} (1-\alpha)^k C(\mathbf{f}_{t-k}^T),
    \label{eqn:Signal}
\end{align}
where $\alpha\in[0,1]$. Hence, the model keeps memory of its past, but the memory is suppressed exponentially. Each day $t$, the platform operator updates the model in Eqn.~\eqref{eqn:Signal} and makes it accessible to his subscribers $\mathcal{T}$ as a travel time performance signal in the decision support tool [compare Eqn.~\eqref{eqn:DecisionSupportDynamics}]. 

\section{Results}

In this section, we characterize the collective dynamics of urban commuting with crowdsourced route choice support introduced in the previous section. First, we illustrate how crowdsourced decision support yields a separation of commuter flows into technology and non-technology users for strongly opportunistic route choice behavior. Second, we demonstrate how this flow splitting fuels systemic inefficiencies. Third, we unveil how the collective sharing of traffic information may stimulate congestion and leave certain routes unused for long timescales. 

\subsection{Pigou example}

In the following, we analyze a minimal model street network $G$ consisting of a single OD-pair whose origin and destination nodes ($O$ and $D$) are connected via two routes 1 and 2 (see Fig.~\ref{fig:FIG3_PigouSystem}b). In this system, commuters draw from homogeneous strategy profiles $\mathcal{S}_i=\{1,2\}$ for all $i\in\mathcal{N}$. Let the latency structure on $G$ be given by
\begin{align}
    c_1(f_1) = \left(\frac{f_1}{f_0}\right)^4,\qquad c_2(f_2) = 1,
\end{align}
where $f_0\in\mathbb{R}^+$ controls the flow capacity along route 1 (see section \ref{sec:UrbanEnvironment}). As shown by Pigou and Roughgarden the combination of fixed and variable latency routing options creates strong incentives for road users to travel along route 1 if $f_0$ is sufficiently large, but does not settle into a socially optimal user equilibrium \cite[Ch. VIII \S 4]{Pigou1920}\cite{Roughgarden2002a}. On this network define the adaptive payoff-based learning dynamics with crowdsourced traffic information specified in Eqn.~\eqref{eqn:DecisionSupportDynamics}.

In Fig.~\ref{fig:FIG3_PigouSystem} we illustrate the resulting commuting dynamics for a population of $N=1000$ people which of a subset $\mathcal{T}\subset \mathcal{N}$ subscribes to the decision support service. We contrast both situations where $\mathcal{T}=\emptyset$ (Fig.~\ref{fig:FIG3_PigouSystem} top row) and $\mathcal{T}\neq \emptyset$ (Fig.~\ref{fig:FIG3_PigouSystem} bottom row). For $\mathcal{T}\neq \emptyset$, we assume homogeneous $\kappa_i=0.5$ for all $i\in\mathcal{T}$ and $\kappa_i = 0$ for all $i\in\mathcal{A}$. Moreover, we assume homogeneous route choice preferences $\beta_i\equiv \beta$ for all $i \in\mathcal{N}$.

\begin{obs}[Separation of flows] The presence of crowdsourced digital route choice support yields a separation of flows into technology users and non-technology users across the two routing options, depending on the route choice preferences $\beta$.
\end{obs}

In Fig.~\ref{fig:FIG3_PigouSystem}, we show the relative flow time series $\{f_{1,t}^X/X\}_{t=1}^M,\ M\in\mathbb{N}$ along route 1 normalized by group size $X$, namely $X=T$ for technology users and $X=A$ for non-technology users. We contrast the relative flow evolution with and without decision support for $\beta=1.5$ and $\beta=4.5$.

For $\mathcal{T}=\emptyset$ and $\beta=1.5$ the commuter flow along route 1 relaxes to a value slightly below the threshold capacity $f_0$ (see Fig.~\ref{fig:FIG3_PigouSystem}a, dashed) and is subject to minor fluctuations. As $\beta$ increases (see Fig.~\ref{fig:FIG3_PigouSystem}c) the fluctuations increase significantly (see Supplementary Material for details), whereas the long-time average (denoted in terms of a moving average) hints at stationary expectation values for both regimes of route choice preferences. Commuters evolve into a state where they use up all capacity $f_0$ along route 1 up to some buffer that absorbs natural fluctuations. That way, commuters avoid overloading route 1 which would become congested otherwise.

For $\mathcal{T}\neq \emptyset$ we assume $T=f_0=700$ and homogeneous technology trust $\kappa=0.5$ for technology users. Fig.~\ref{fig:FIG3_PigouSystem}d shows even contribution to the relative flow along route 1 from both commuter groups for $\beta=1.5$:
\begin{align}
    \langle f_1^T\rangle /T \approx \langle f_1^A \rangle /A  
\end{align}
Crowdsourced route choice support does not alter the dynamics observed for $\mathcal{T}=\emptyset$. 

For $\beta=4.5$, however, flows evolve qualitatively different with and without technology support. The relative flow contribution of the two commuter groups separates. $\mathcal{T}$ has a significantly higher relative contribution to $f_{1,t}$ compared to $\mathcal{A}$ (see Fig.~\ref{fig:FIG3_PigouSystem}f):
\begin{align}
    \langle f_1^T\rangle /T > \langle f_1^A \rangle /A    
\end{align}
Hence, technology users make use of the majority of the available capacity along the street segment and displace non-technology users onto the fixed latency route 2. The phenomenon persists and lasts beyond the transient. 

Fig.~\ref{fig:FIG3_PigouSystem}e demonstrates that flow separation emerges upon exceeding a critical value of $\beta$ and intensifies the more opportunistic the population's route choice behavior is. Simulations indicate that the critical value of $\beta$ is always close to two, but is modulated by the parameters of the model (see Supplementary Material for details).

\begin{figure*}
    \centering
    \includegraphics[width=\linewidth]{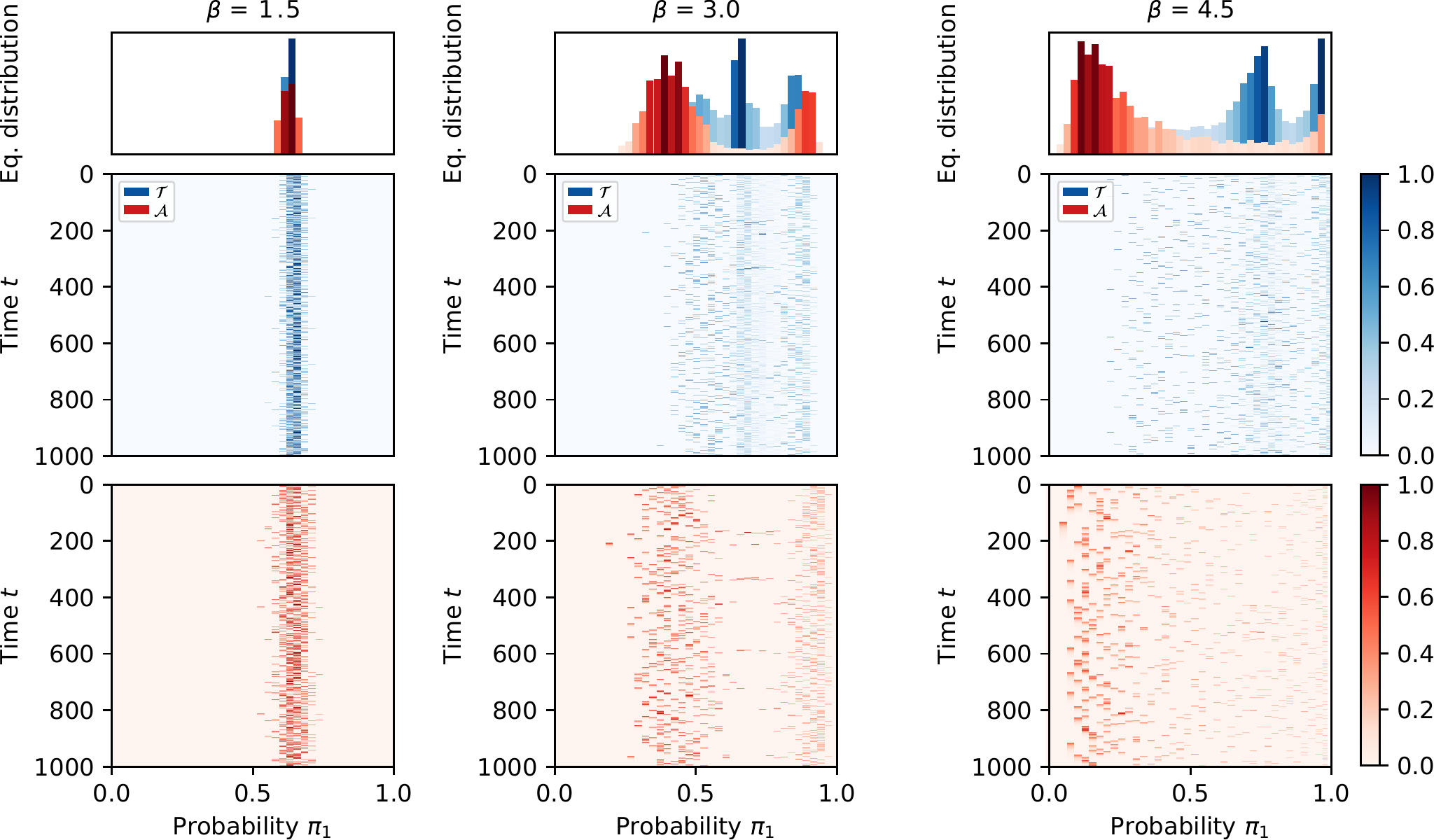}
    \caption{\textbf{Heterogeneous mixed strategy profiles give rise to flow separation}. For $\beta>\beta_\mathrm{crit.}$ the distribution of mixed strategy profiles becomes heterogeneous and the probability mass separates for technology and non-technology users. The mixed strategy profiles of technology users are described by a bimodal distribution that exhibits peaks at $\pi_1=f_0/N$ and $\pi_1=1$. Thus, technology users learn to preferably choose route 1 while non-technology users evolve into selecting the route alternative. Simulation parameters are $N=1000, T=700, f_0=700, \alpha=0.5, \kappa=0.5$.}
    \label{fig:FIG4_MixedStrategyHistogram}
\end{figure*}

\begin{obs}[Heterogeneity in mixed strategy profiles] Flow separation emerges from heterogeneously distributed mixed strategy profiles between technology and non-technology users.
\end{obs}

Fig.~\ref{fig:FIG4_MixedStrategyHistogram} shows the distribution of mixed strategy profiles in the commuter population of technology and non-technology users. 

For $\beta=1.5$ all commuters' individual mixed strategy profiles cluster at $\pi_1=(f_0-\epsilon)/N$ (where $\epsilon>0$ corresponds to the spare capacity that commuters leave to absorb natural route choice fluctuations), independent of technology use or not, explaining the long-time averages in Fig.~\ref{fig:FIG3_PigouSystem}d that fluctuate around this value. Qualitatively, technology and non-technology users behave the same and can be approximated identical in a meanfield description (see Fig.~\ref{fig:FIG4_MixedStrategyHistogram}a). 

As $\beta$ increases beyond a value of approximately two for the given configuration (see Fig.~\ref{fig:FIG4_MixedStrategyHistogram}e, blue shading), the distribution of mixed strategy profiles becomes heterogeneous for technology and non-technology users (see Fig.~\ref{fig:FIG4_MixedStrategyHistogram}b). For technology users the distribution shifts in favor of route 1, whereas it shifts towards route 2 for non-technology users (see Fig.~\ref{fig:FIG4_MixedStrategyHistogram}c). Technology users learn to preferably use the variable latency route 1, which has sufficient capacity $f_0=T$ to cater for $\mathcal{T}\subset\mathcal{N}$, while non-technology users are displaced onto the fixed latency route 2. With crowdsourced traffic information the distribution of mixed strategy profiles becomes bimodal in this parameter constellation. It exhibits peaks at $\pi_1=1$ and $\pi_1=f_0/N$. Hence, any meanfield-type of approximation breaks down in this regime of route choice preferences. 

This dynamics is a result of the information asymmetry between the two commuter groups: In every time step technology users obtain traffic information about the different routing options, while non-technology users need to sample the information on their own route by route, resulting in different effective learning rates. Technology users achieve an effective first-mover advantage and pre-occupy the seemingly low latency route 1 while non-technology users are still sampling the traffic observations to decide between the different routes. Consequently, non-technology users realize that route 1 is already occupied to capacity and becomes congested if they join in as well. Hence, their rational choice is to select route 2 (see Supplementary Material for details).

The presence of digital routing technology effectively turns the congestion game into a sequential move game where technology users inherit the first mover advantage of the digital platform operator, acting as a Stackelberg leader. Flow separation is its macroscopic manifestation and is rooted in heterogeneous travel time beliefs between the two commuter groups.

\begin{obs}[Externalities from crowdsourcing] Crowdsourced route choice support does not reduce the average travel time for the population, but may cause its rise.
\end{obs}

\begin{figure*}
    \centering
    \includegraphics[width=\linewidth]{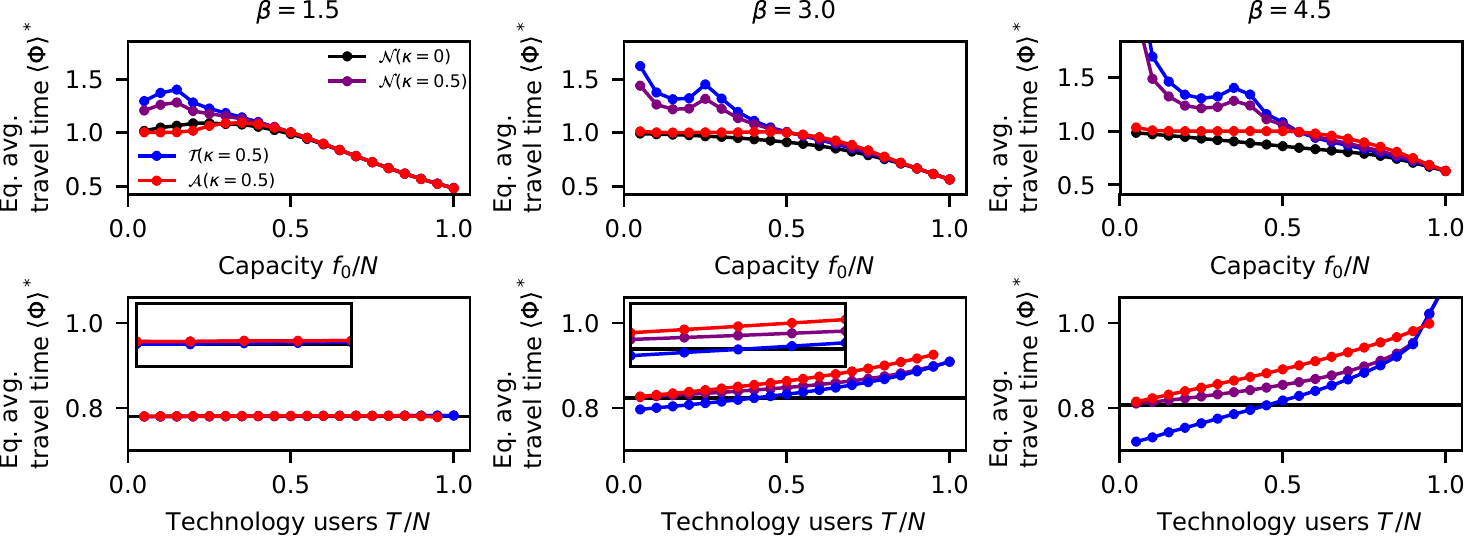}
    \caption{\textbf{Impact of crowdsourced traffic information on average travel time}. (top row) In situations where the number of technology users outnumbers the available street capacity ($f_0\ll T$) the presence of crowdsourced digital decision support leads to higher average commuting times compared to the situation without technology support ($\kappa=0$). Shown here for $T/N=0.7$. (bottom) For given street capacity $f_0/N=0.7$ the average commuting time rises in the number of technology users and $\beta$, and is always higher than for $\kappa=0$. Technology users may realize shorter travel times compared to their non-technology enabled fellow travelers who suffer longer average trip durations. Crowdsourced digital decision support imposes externalities.
    Simulation parameters are $N=1000$ and $\alpha=0.5$.}
    \label{fig:FIG5_PhasePlots}
\end{figure*}

Fig.~\ref{fig:FIG5_PhasePlots} illustrates the impact of crowdsourced traffic information on the equilibrium average commuting time per person (purple) and contrasts the results to the situation where no digital technology were present (black). 

In all $\beta$-regimes the average travel time responds qualitatively the same to a change in the latency structure of the street network (see Fig.~\ref{fig:FIG5_PhasePlots} (top row)). Upon expanding the capacity $f_0$ on route 1 travel times generally decrease in the absence of digital technology. However, assuming a fixed number of commuters to use route choice apps ($T/N=0.7$) the average travel time is always higher than in the corresponding scenario where $T=0$. In all $\beta$-regimes the travel time rises substantially in situations where $f_0/T\ll 1$ and approaches the equilibrium values for the situation without technology if $f_0/T\gg 1$. Moreover, extending the capacity $f_0$ on route 1 when it is small, yields a temporary increase of the average travel times under digital route choice support, before travel times decrease as the capacity exceeds the number of technology users. Crowdsourced traffic information does not reduce congestion in these constellations, but instead intensifies it.

Fig.~\ref{fig:FIG5_PhasePlots} (bottom row) illustrates how the average travel time per person depends on the share of technology users for fixed street latency structure ($f_0/N=0.7$). In the $\beta$-regime where there is no flow separation, the average travel time is independent of the share of technology users in the population (see Fig.~\ref{fig:FIG5_PhasePlots} bottom left). However, as commuter flows separate for more opportunistic route choice preferences the average travel time increases monotonously in the number of technology users (see Fig.~\ref{fig:FIG5_PhasePlots} bottom center and right). It is always higher than in the situation without decision support. If $T\ll f_0$, technology users realize a travel time advantage compared to non-technology users in the given configuration. As $T\approx f_0$ or $T\gg f_0$ both commuter groups face higher average travel times compared to the situation where $T=0$. Consequently, crowdsourced decision support never improves the average travel time for the population. Only subsets of commuters benefit from the crowdsourced traffic information, realized at the expense of the remainder of the population. The average travel time increment for the total population defines an externality from crowdsourcing and fuels systemic inefficiencies (see Supplementary Material).

\begin{figure*}
    \centering
    \includegraphics[width=\linewidth]{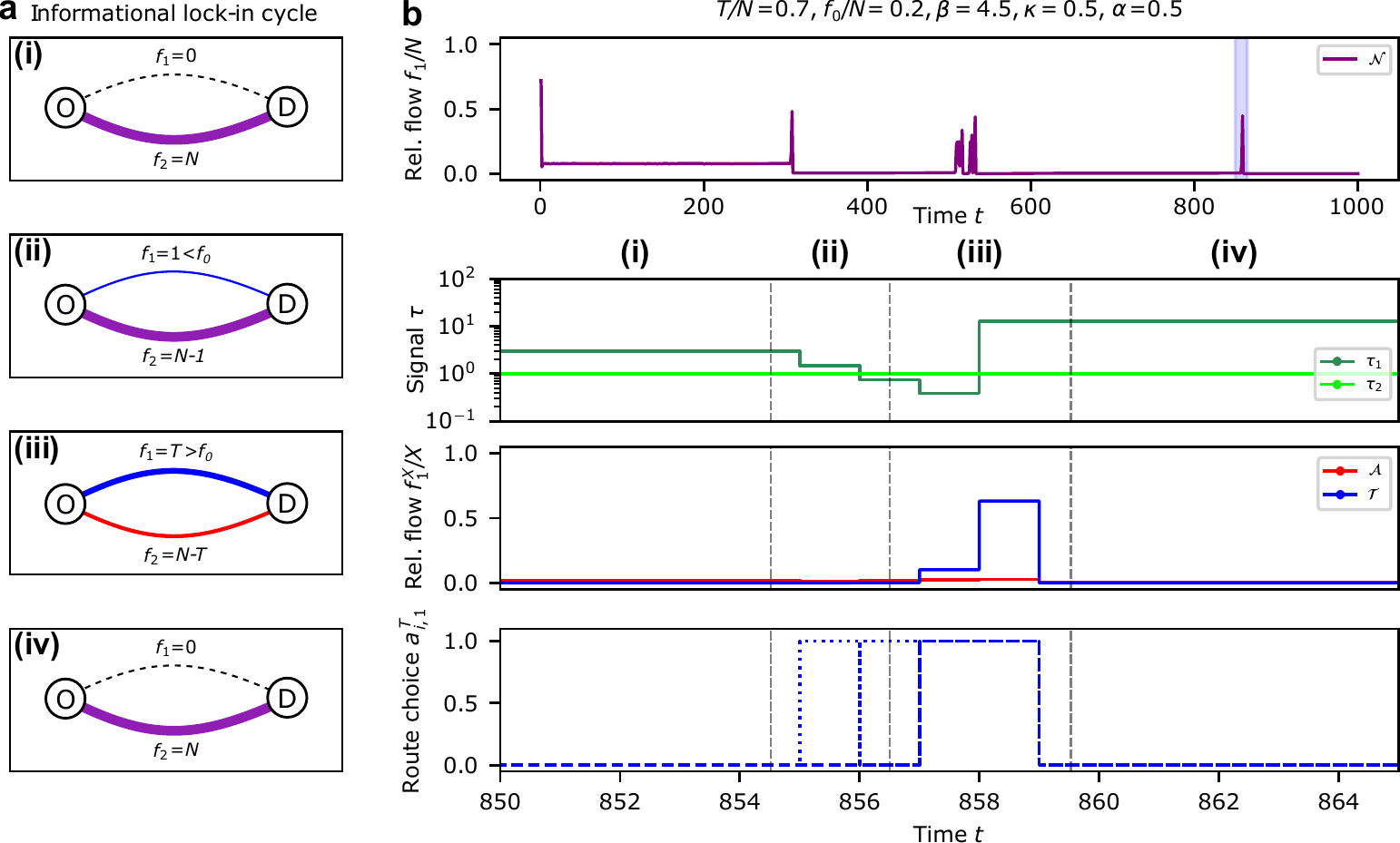}
    \caption{\textbf{Sharing information prevents use of low latency routes}. \textbf{a}, In an informational lock-in cycle technology users cannot unilaterally switch to low latency routes, illustrated here for the Pigou network. In a four step cycle commuters are forced to switch back to a route alternative. (i) If the digital platform provider cannot crowdsource up-to-date traffic observations he provides outdated signals, potentially deterring technology users to choose the corresponding route. (ii) A single individual choosing the route by chance provides the required flow information and reveals that a route alternative yields attractive travel time. (iii) The individual's discovery is shared among all technology users and induces collective action. Technology users will switch from route 2 to route 1 in unison. (iv) Route 1 becomes congested upon the sudden flow spike. The platform provider adapts his signal, discourages the use of route 1, and induces a collective back-switch to route 2. \textbf{b}, In the simulation the informational lock-in cycle emerges for $f_0/T\ll 1$ and large $\beta$. At the time corresponding to regime (i) route 1 is unused, but technology users receive a signal $\tau_1>1$, discouraging to switch. At stage (ii) a single individual $i\subset\mathcal{T}$ chooses route 1, causing an adaptation of $\tau_1$ which remains larger than $\tau_2$ owing to the learning rate $\alpha=1/2$ of the platform provider. Another single individual $j\subset\mathcal{T}$ provides a further update on the flow along route 1 in the next time step, leading to $\tau_1<\tau_2$. The next day approx. 70$\%$ of technology users collectively switch to route 1 which becomes heavily congested. $\tau_1$ rises substantially and technology users learn to avoid route 1. However, even non-technology users may learn to avoid the potentially faster route 1 if they experience the heavy flow spike induced by technology users, eventually leaving route 1 unused instead of being occupied by non-technology users.}
    \label{fig:FIG6_InformationalLockIn}
\end{figure*}

\begin{obs}[Informational lock-in cycle] Crowdsourced traffic information may prevent technology users from unilaterally switching to low latency routes and leave route capacity unused.
\end{obs}

Fig.~\ref{fig:FIG5_PhasePlots} (top row) illustrates how technology users may be subject to more congestion compared to their non-technology using fellow travelers in constellations where $f_0/T\ll 1$. This inversion of travel time advantages results from an informational lock-in effect: The collective sharing of traffic information within the group of technology users prevents unilateral switching to low latency routes. Any discovery of faster routing options induces collective action from all technology users, even if the group size exceeds the route's maximum capacity. Formerly faster routing options become strongly congested, forcing the commuter population to switch back to their route alternatives. Technology users face a four-step informational lock-in cycle (see Fig.~\ref{fig:FIG6_InformationalLockIn}a):
\begin{enumerate}
    \item \textit{Emergence of low latency route}: A route connecting origin and destination is underutilized and enables faster travel time than the alternatives.
    \item \textit{Discovery of route}: A technology user discovers the low latency route and automatically shares the information with the digital platform provider. The platform provider updates his travel time signals and shares them with all of his subscribers.
    \item \textit{Collective switch}: Technology users collectively switch to the low latency route based on the signal received from the platform provider. The route becomes strongly congested as it is overutilized. 
    \item \textit{Collective backlash}: Technology users learn to avoid the route. The platform provider adapts his travel time signals and discourages its use. In the next game stage technology users will collectively switch back to their routing alternatives.
\end{enumerate}
In Fig.~\ref{fig:FIG6_InformationalLockIn}b we demonstrate the informational lock-in cycle for the Pigou system with crowdsourced decision support for $f_0/T=2/7<1$ and $\beta=4.5$. Commuters experience the cycle repeatedly on timescales of few hundred timesteps, eventually clearing route 1 completely independent of technology or non-technology users. The collective phenomenon repeats on corresponding timescales and adds spikes to the average travel time of the population. Consequently, the population's average travel time rises beyond a value of one (see Fig.~\ref{fig:FIG5_PhasePlots} (top row)), even though the population chooses the fixed latency route 2 almost exclusively.\\

\section{Summary} 

Crowdsourced digital decision support can induce different forms of collective action among technology users, inevitably impacting the travel behavior of non-technology users, too. In the regime of strongly opportunistic route choice behavior crowdsourced digital decision support fuels selfish route choices and causes flow splitting -- the separation of technology and non-technology users across routes. Depending on the number of technology users within the population and the latency structure of the street network digital decision support causes travel time externalities. Either technology users realize travel time advantages at the expense of non-technology users, or they face a travel time disadvantage resulting from an informational lock-in effect. In both cases, the overall travel time of the population can only increase.

\section{Conclusion and discussion}

Digital routing technology has become a staple of urban commuting. In this article, we demonstrated how smartphone-based route choice support tools impact the collective traffic assignment in urban commuting games under uncertainty. 

By modifying congestion games under adaptive payoff-based learning dynamics to cater for digital decision support, we illustrated how digital travel time performance signals alter commuters' route choice rationale, and may thus shift macroscopic traffic flows. For collective sharing of crowdsourced traffic observations heterogeneous travel time beliefs emerge under certain conditions and give rise to a separation of technology and non-technology enabled commuter flows along different routes. As we show, collective sharing of crowdsourced route choice support may promote systemic inefficiencies and leave street capacity unused instead of removing the externalities long-term.

To isolate the interactions induced by digital route choice support, the above analysis is based on a minimal model of day-to-day commuting dynamics. We expect similar interactions in larger networks, as the observations are driven by the different latency functions between the fastest route and the other route options. Technology users naturally gravitate towards the route options with lowest latency and will either displace other commuters (flow splitting) or enter an informational lock-in cycle depending on the capacity of the fastest route. On shorter timescales, dynamic adaptation of the route choices and suggestions during trips may introduce additional effects and may contribute to an explanation for spontaneous formation of congestion \cite{Szeto2006,Jiang2019}, for example depending on the type and limits of the available information.

Despite the potential negative effects of crowdsourced traffic information, alternative protocols of digital route choice support, aimed at reducing systemic inefficiencies, may reallocate traffic flows towards a socially preferable allocation across routes as well as modes of transportation. Thereby, mode-route choice support tools become powerful demand management instruments that enable central coordination of microscopic routing decisions, helping, for example, to alleviate the negative socio-economic effects of congestion-induced delays in growing cities as urbanization accelerates \cite{Depersin2017,Louf2015}. Routing applications have the potential to information-theoretically realize Stackelberg routing schemes as originally proposed in Refs.~\cite{Korilis1997, Roughgarden2002a}. Finding suitable algorithms and protocols that result in socially preferable flow allocations points at important future research directions. New types of route choice support tools will not only become a fundamental part of Mobility-as-a-Service platforms, but hold the potential to minimize negative social, economic and environmental impacts of congestion in urban commuting.

\bibliography{FlowSplitting}

\section*{Acknowledgements}
D.S. acknowledges support from the Studienstiftung des Deutschen Volkes. M.T. acknowledges support from the German Research Foundation (Deutsche Forschungsgemeinschaft, DFG) through the Cluster of Excellence Center for Advancing Electronics Dresden (cfaed).

\section*{Competing interest}
The authors declare no competing interests.

\section*{Correspondence}
Correspondence and requests should be addressed to D.S. and M.T. 

\section*{Data availability}
The data that support the findings of this study are available from the  authors upon reasonable request.

\section*{Supplementary Material}
The Supplementary Material contains additional simulations and analyses on the impact of crowdsourced traffic information in technology-enabled commuting games, as well as details on methods applied to derive the results in this Main Manuscript.

\end{document}